\documentclass{nature}
\usepackage{amssymb,graphicx}

\title{Predictability of El Ni\~no as a Nonlinear Stochastic Limit Cycle} 

\author{Debabrata Panja$^{1}$ and Gerrit Burgers$^{1}$}

\begin{document}

\maketitle

\begin{affiliations}
 \item Royal Netherlands Meteorological Institute (KNMI), Postbus 201,
 3730 AE De Bilt, The Netherlands
\end{affiliations}

\begin{abstract}
Abstract: The El Ni\~no phenomenon, synonymously El Ni\~no-Southern
Oscillation (ENSO), is an anomalous climatic oscillation in the
Equatorial Pacific that occurs once every 3-8 years. It affects the
earth's climate on a global scale. Whether it is a cyclic or a
sporadic event, or whether its apparently random behaviour can be
explained by stochastic dynamics have remained matters of
debate. Herein ENSO is viewed, unconventionally, as a two-dimensional
dynamical system on a desktop. The main features of ENSO:
irregularity, interannual variability, and the asymmetry between El
Ni\~no and La Ni\~na are captured, simply, comprehensibly, quickly and
cheaply, in a nonlinear stochastic limit cycle paradigm. Its
predictability for ENSO compares remarkably well with that of the best
state-of-the-art complex models from the European and American
meteorological centres. Additionally, for the first time, by analyzing
subsurface Equatorial Pacific data since 1960, this model finds that
long-term variations are not caused by ENSO itself, but by external
sources.

\end{abstract}

El Ni\~no is an anomalous climatic event in the Equatorial Pacific. On
average, the western Pacific sea-surface temperature (SST) is about
$4^\circ$-$5^\circ$C higher than that of the eastern Pacific, and the
thermocline, or the sharp vertical temperature gradient separating the
warm and the cold water, lies deeper in the western Pacific than in
the Eastern Pacific. The higher SST of the western Pacific results in
rising warm, moist air from the Indonesian coast, and the descent of
cool, dry air towards the eastern south Pacific. These atmospheric
movements are supplemented by the high altitude eastward jet stream
and the westward trade winds along the Equatorial Pacific sea
surface. During an El Ni\~no, which peaks in the northern winter, the
eastern Pacific gets warmer as the pool of warm water stretches much
further eastward. The western Pacific thermocline becomes shallower
and the westward trade winds at the sea surface become much weaker,
and sometimes even reverse. These anomaly conditions in the Equatorial
Pacific affect the earth's climate at a global scale.

That ENSO is a result of ocean-atmosphere interaction has been clear
since the time of Bjerknes\cite{bjerknes}. It has been
well-established that there are three key players in the
ocean-atmosphere interaction that determine the ENSO dynamics at
interannual time-scales: the SST, the atmospheric wind stress anomaly
and the ocean's ability to transport energy\cite{philander} in the
Equatorial Pacific. However, how the dynamics of these quantities
relate to each other to perturb the normal climatic conditions to
create an El Ni\~no, and especially how the growth of an El Ni\~no is
controlled so that the normalcy is restored have remained matters of
discussion.\cite{neelinrev} A large number of models geared to address
these issues --- ranging from dynamical models inspired by the physics
of
ENSO\cite{philanderpap,hirst,cane1,zebiak,battistialone,suarez,graham,battisti,neelin1,wang,jin3,vaart,neelin2}
to stochastic
ones\cite{philander2,lau,burgers1,neelin,jin0,eckert,kleeman,tziperman,flugel,penland,tahl}
--- have put forward different mechanisms at varying degrees of
complexity: for some ENSO is inherently
unstable\cite{philanderpap,hirst} and the nonlinearities in ENSO
dynamics eventually control the growth of an El
Ni\~no\cite{cane1,zebiak,battistialone,suarez,graham,battisti}, while
some others have assumed that the normal climatic condition is stable
and that ENSO is excited by
noise\cite{wang,philander2,jin3,lau,neelin,eckert,kleeman,tziperman,flugel,penland,tahl}.
These models have also differed in their interpretation of the role of
noise originating from external processes at a faster time scale, and
the role of nonlinearity in ENSO properties. Over and above, they can
capture different features of ENSO in their own ways; nevertheless,
the ability to reproduce all the basic features of ENSO, while  being
able to simultaneously predict ENSO with good accuracy in a
comprehensive manner is still lacking in the ENSO model
world\cite{neelinrev}. Remarkably, one such feature of ENSO not easily
reproduced is the skewness of the ENSO indices, or the asymmetry
between El Ni\~no and La Ni\~na\cite{steph}.

Given this setting, herein ENSO is viewed, unconventionally, as a
two-dimensional dynamical system on a desktop, described by the
Ni\~no3 index, the anomaly in the average SST of the region
$5^\circ$S-$5^\circ$N, $150^\circ$W-$90^\circ$W, and the anomaly in
Z$20$ or the average depth of the $20^\circ$C isotherm (as a proxy for
the thermocline depth) of the region $5^\circ$S-$5^\circ$N,
$120^\circ$E-$290^\circ$E, henceforth denoted by $T$ and $H$
respectively. Using the Ni\~no3 and Z20 data averaged over month $i$,
the (discrete) ENSO dynamical system indexed by $(T_i,H_i)$ is
constructed. Based on the physics of ENSO, a phenomenological model,
subject to fixed-amplitude Gaussian white noise, is conjectured to
describe this dynamical system. The model captures, simply,
comprehensibly, quickly and cheaply, all the main features of ENSO:
irregularity, interannual variability, and the asymmetry between El
Ni\~no and La Ni\~na, in a nonlinear stochastic limit cycle
paradigm. Its predictability for ENSO compares remarkably well with
that of the best state-of-the-art complex models from the European and
American meteorological centres. Additionally, for the first time, by
analyzing  subsurface Equatorial Pacific (Z20) data since 1960
(roughly the time when regular record-keeping of subsurface Pacific
temperature profile began), this model finds that decadal (and above)
variations stem from external sources, e.g. climate shifts.

\section*{ENSO as a nonlinear stochastic limit cycle}   

The idea behind viewing ENSO as a two-dimensional dynamical system is
based on the so-called ``recharge
oscillator''\cite{jin1,jin2,jin3}, which identifies the
Equatorial eastern Pacific SST anomaly $T_E$, the equatorial eastern
and the western Pacific thermocline depth anomalies $H_E$ and $H_W$,
and the central Pacific zonal wind-stress anomaly $\tau$ as the main
oceanic and atmospheric quantities involved in the ENSO dynamics. The
Equatorial eastern and the western Pacific thermocline depth anomalies
can be combined to form two independent variables: $(H_E-H_W)$,
representing the ``thermocline anomaly tilt'', and $(H_E+H_W)$,
representing the anomaly in the amount of warm water volume  present
in the Equatorial Pacific
(WWVA)\cite{jin1,jin2,jin3,kessler,meinen2}. Of these two, a positive
anomaly in $(H_E-H_W)$ is strongly in phase with a positive eastern
Pacific SST anomaly and a positive central Pacific zonal wind-stress
anomaly. As envisaged by the recharge oscillator, before the onset of
an El Ni\~no, (meridional) Sverdrup transport of warm water towards
the Equator gives rise to a positive anomaly in the WWVA. A positive
WWVA gives rise to a positive $T_E$. Subsequently, $\tau$ responds
positively to a positive $T_E$, and increases $(H_E-H_W)$. These
changes in the Pacific surface conditions then generate Rossby and
Kelvin thermocline waves in the equatorial waveguide: these waves
first make the western Pacific thermoclines shallower, and the return
of the reflected waves at the western Pacific boundaries, further on,
reduces the eastern Pacific SST anomaly.

The recharge oscillator does not explain how the (meridional) Sverdrup
transport of warm water towards (or away from) the Equator is
triggered. Nonetheless, it seems logical that a positive WWVA results
in a positive $T_E$ and vice versa. A linear relationship between
these two quantities was conjectured by Burgers {\it et
al.}\cite{jin3}, but that is in contradiction with the observation
data: the magnitude of the positive $T_E$ due to a given positive
magnitude of the WWVA during an El Ni\~no is larger than the magnitude
of $T_E$ due to a negative WWVA of the same magnitude during a La
Ni\~na\cite{meinen2,mcphadden}. This asymmetry between the El Ni\~no
and the La Ni\~na (manifested equivalently via the Ni\~no3
skewness\cite{steph}), not fully understood at present, indicate that
how the WWVA and $T_E$ interact is unclear: scenarios based on air-sea
fluxes feedback on the SST, and the oceanic upwelling and vertical
mixing mechanism have been suspected to contribute to
it\cite{wang1,wang2}. The fact however remains that the WWVA leads
$(H_E-H_W)$, $T_E$ and $\tau$, which are strongly in phase with each
other, by approximately 7 months on average, and consequently it
should be considered as an important predictor for  $T_E$ at ENSO
time-scales. The natural variables in the (minimalistic) model for
ENSO herein, therefore, are the Ni\~no3 monthly anomaly as a
representative of $T_E$, and the Z20 monthly anomaly, representing the
WWVA (hereafter denoted by $T$ and $H$ respectively), forming a
two-dimensional dynamical system. The lack of our understanding of how
the WWVA and $T_E$ interact motivated this study of ENSO as a
stochastic dynamical system, based on empirical data.

\begin{figure}[h]
\begin{center} 
\begin{minipage}{0.48\linewidth}
\includegraphics [width=0.8\linewidth,angle=270]{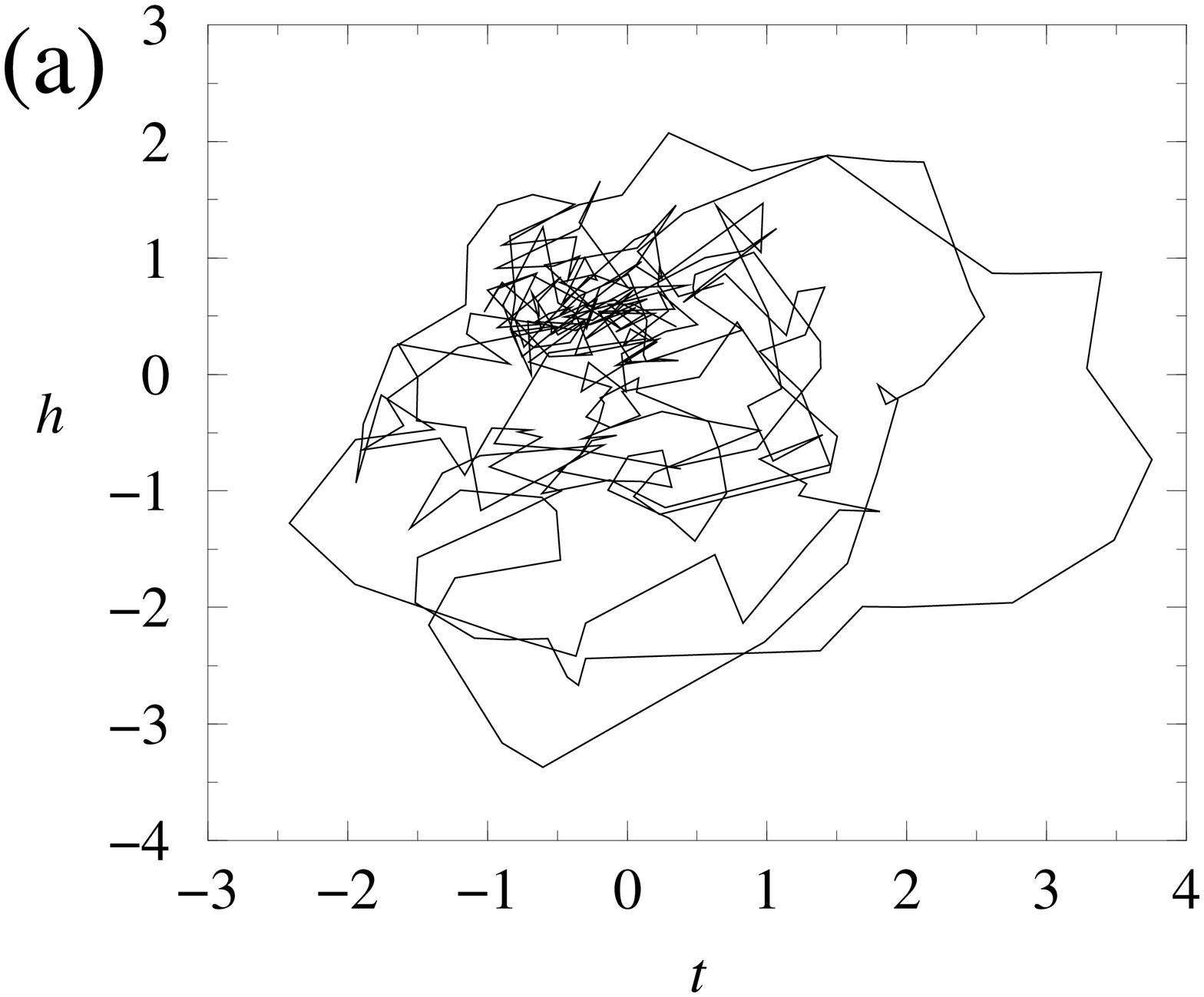}
\end{minipage}
\begin{minipage}{0.48\linewidth}
\includegraphics [width=0.8\linewidth,angle=270]{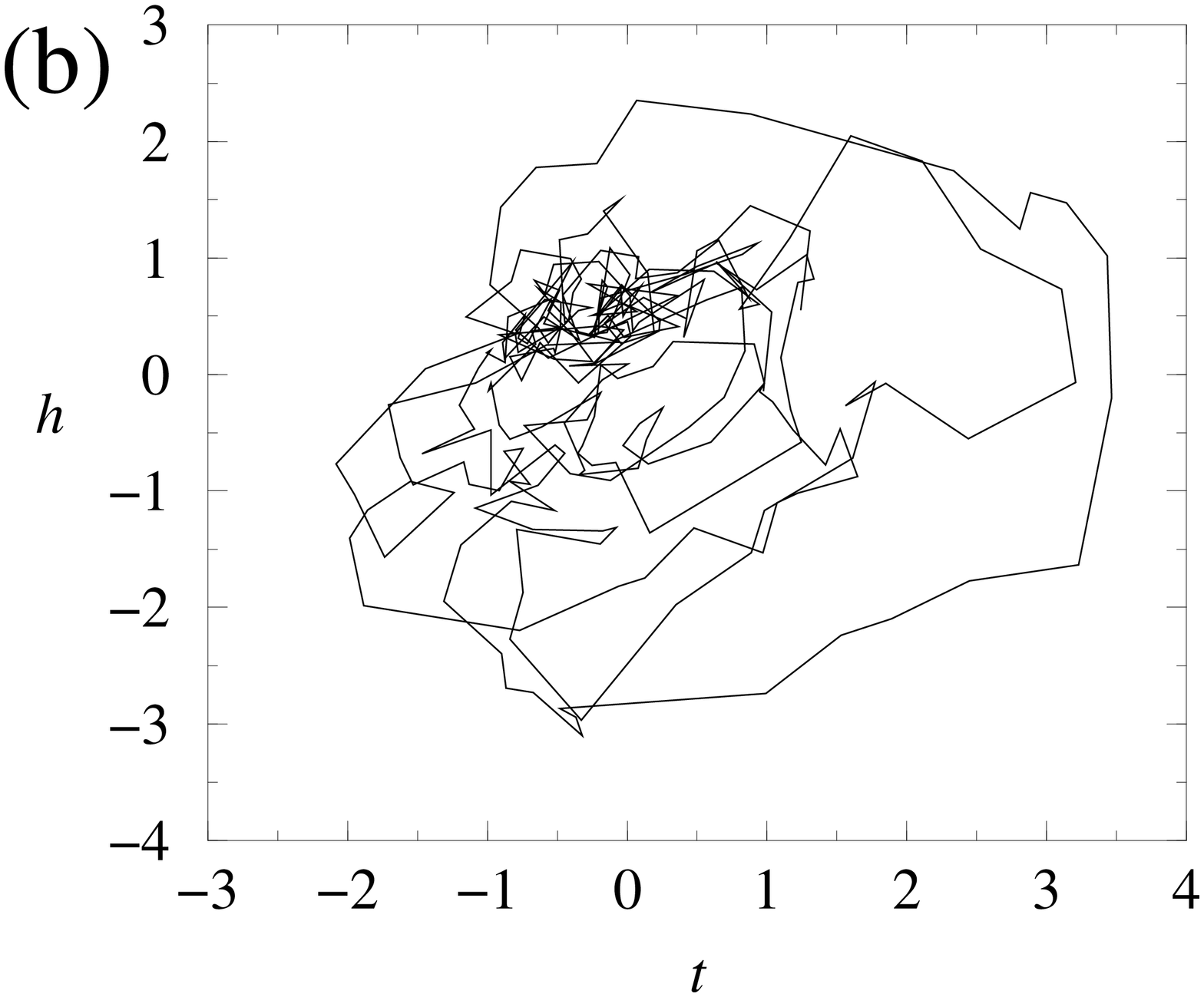}
\end{minipage}
\caption{(a) Normalized monthly anomaly data for Ni\~no3 ($t$) and Z20
($h$) in the Equatorial Pacific, describing ENSO for 1980-2004
(source: MET office objective analysis for the ENSEMBLES project,
hereafter referred to as the ``ENSEMBLES data''; publicly available at
http://www.ecmwf.int/research/EU\_projects/ENACT/ocean\_analyses/index.html);
(b) The same for the 1980-2002 data obtained from 
Australian Bureau of Meteorology (hereafter ABOM); publicly available at
http://www.bom.gov.au/bmrc/ocean/results/climocan.htm\#subsurface\%20anals].
\label {fig1}}
\end{center}
\end{figure}

The monthly Ni\~no3 and Z20 anomalies in dimensionless units (rendered
dimensionless by normalizing w.r.t. their r.m.s. magnitudes),
hereafter denoted by $(t,h)$, are shown in Fig. \ref{fig1}. A positive
$t$ (resp. $h$) means a positive Ni\~no3 anomaly (resp. WWVA) and vice
versa. In this notation, the aim of this study is to describe the ENSO
dynamics stochastically as
$t_{i+1}=f(t_i,h_i)+\xi_i,h_{i+1}=g(t_i,h_i)+\eta_i$, where $f(x,y)$
and $g(x,y)$ are two nonlinear functions of their arguments, and $\xi$
and $\eta$ are mutually uncorrelated fixed-amplitude Gaussian white
noise. Visual inspection of the data in Fig. \ref{fig1} suggests
strong locking of the ENSO dynamics to its phase [defined by
$\phi=-\tan^{-1}(h/t)$], and therefore a cylindrical co-ordinate
system $r=\sqrt{t^2+h^2}$ and $\phi$, is more suitable. In these
co-ordinates, without any loss of generality, the nonlinearities in
the ENSO dynamics are then expressed by two coupled nonlinear
differential equations as: $dr/dt\equiv r\gamma(r,\phi)$ and
$d\phi/dt\equiv\omega(r,\phi)$, with the $(r_i,\phi_i)$-values
obtained from the corresponding $(t_i,h_i)$ ones.

Based on the resemblance of Fig. \ref{fig1} to a limit cycle (i.e.,  a
tendency to grow when the system is close to $r=0$, combined with a
tendency to decay when the system grows far away form $r=0$)  strongly
locked to its phase, the ansatz
$\gamma(r,\phi)=f_1(\phi)[f_2(\phi)-r]$ and
$\omega(r,\phi)=g_1(\phi)\sqrt{r}$ is made to describe ENSO. It is
clear from Fig. \ref{fig1} that $f_2(\phi)$ cannot simply be a
constant, and for an explanation for $\omega(r,\phi)\propto\sqrt{r}$
see the methods section. With  this ansatz, the natural choice for
$f_1(\phi)$, $f_2(\phi)$ and $g_1(\phi)$ is clearly a series expansion
in increasing orders of sines and cosine harmonics as
$(A+B_1\cos\phi+B_2\sin\phi+C_1\cos2\phi+C_2\sin2\phi+\ldots)$. The
parameters $A, B_1, B_2,\ldots$ can then be estimated from a
time-series $(t_i,h_i)$. This estimation is performed in the following
manner: the observed time-series for $N$ sequential months were
denoted as $(t_1,h_1)$, $(t_2,h_2)$, $\ldots,(t_N,h_N)$. Then for
$i=1,2\ldots N$, $(t_i,h_i)$ is taken as the starting value and
$dr/dt$ and $d\phi/dt$ are integrated forward from time $i$ to time
$(i+1)$ using the functional forms of $\gamma(r,\phi)$ and
$\omega(r,\phi)$. This yields the model theoretical values $(\tilde
t_{i+1},\tilde h_{i+1})$, as well as the noise ${\mathbf
e}_i=(t_{i+1}-\tilde t_{i+1},h_{i+1}-\tilde h_{i+1})$. The parameters
$A, B_1, B_2,\ldots$ can then be estimated by least-square
optimization, i.e., by minimizing $E=\sum_{i=1}^{N-1}{\mathbf
e}_i\cdot{\mathbf e}_i$. Note that this optimization process not only
yields the parameter values, but also the characteristics of the noise
${\mathbf e}_i$, to be later used to realistically construct $\xi$ and
$\eta$.

\begin{figure}[ht]
\begin{center} 
\begin{minipage}{0.49\linewidth}
\includegraphics [width=0.95\linewidth]{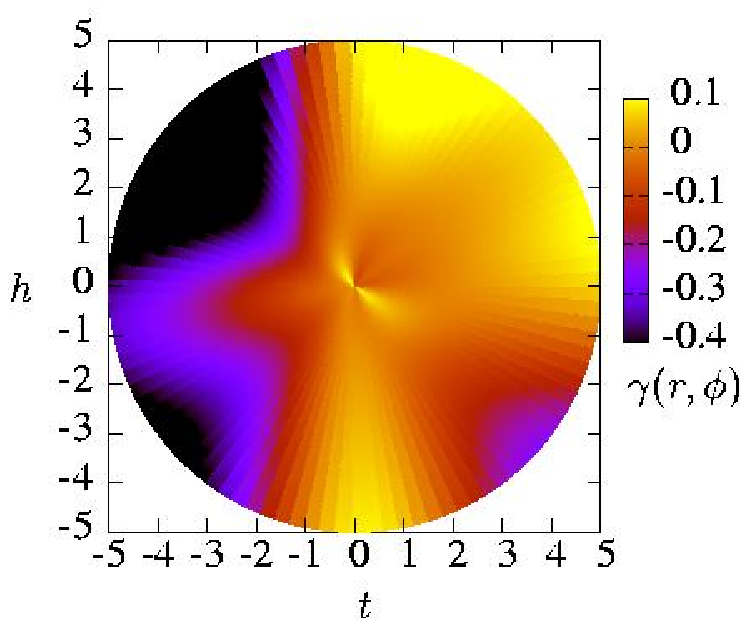}
\end{minipage}
\begin{minipage}{0.49\linewidth}
\includegraphics [width=0.95\linewidth]{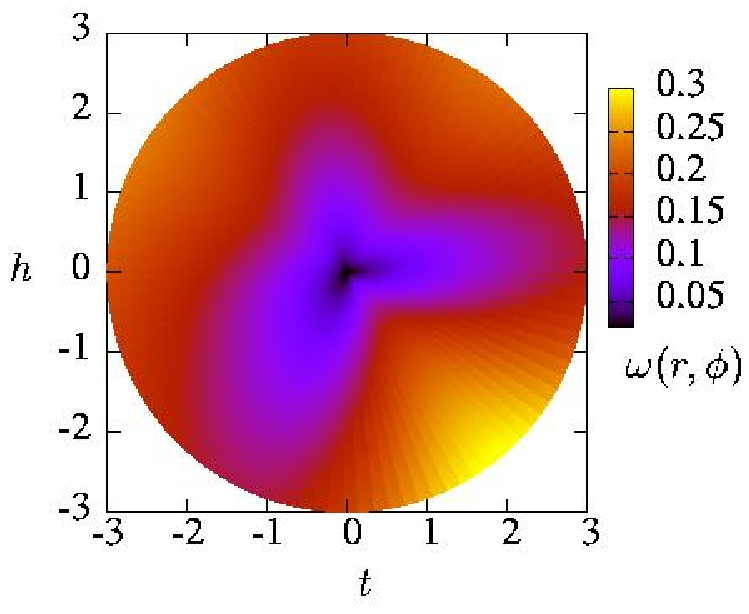}
\end{minipage}
\caption{Phase-space manifold structure for the ENSO dynamics, given
by $dr/dt$ and $d\phi/dt$, in terms of $\gamma(r,\phi)$ and
$\omega(r,\phi)$.The parameters that generated these plots have been
optimized on the 1980-2004 $(t,h)$ ENSEMBLES data
[Fig. \ref{fig1}(a)]. The corresponding phase-space manifold structure
for the ABOM data (not shown here) of Fig. \ref{fig1}(b) is only
marginally different from the above, as it should be; this confirms
the stability of the optimization process to describe ENSO.\label
{fig2} }
\end{center}
\end{figure}
To fix the functional forms for $f_1(\phi)$, $f_2(\phi)$ and $g(\phi)$
that are to be used throughout the entire length of this study, ABOM
data [Fig. \ref{fig1}(b)] were used for the least-square
optimization. To capture the annual variations of $t$ and $h$, any
choice of $f_1(\phi)$, $f_2(\phi)$ and $g(\phi)$ needed to include the
fourth harmonics for the sine and cosine functions (on average, ENSO
has a 4-year period); and simultaneously, due to the finiteness of the
time-series, the number of parameters in $\gamma(\phi)$ and
$\omega(\phi)$ have been kept as low as possible. Eventually, the
functional forms of $f_1(\phi)$, $f_2(\phi)$ and $g(\phi)$ that used
the smallest number of parameters while still leading to the smallest
value of $E$ for the Australian dataset were chosen. This procedure
showed that there are $16$ parameters needed to describe $f_1(\phi)$,
$f_2(\phi)$ and $g(\phi)$ [or $\gamma(r,\phi)$ and
$\omega(r,\phi)$]. See the methods section for details.

As it turns out, the ansatz about the functional forms for
$\gamma(r,\phi)$ and $\omega(r,\phi)$ are very well-chosen. One of its
immediate outcomes is that the noise characteristics that emerged from
Fig. \ref{fig1} can be fairly accurately described as Gaussian
white. The rest of the outcomes: ``stability'' of this method, ENSO
irregularity, variability, Ni\~no3 skewness and predictability are
discussed below.

\section*{Stability of the optimization procedure}

\begin{figure}
\begin{center}
\begin{minipage}{0.49\linewidth}
\includegraphics [width=0.95\linewidth]{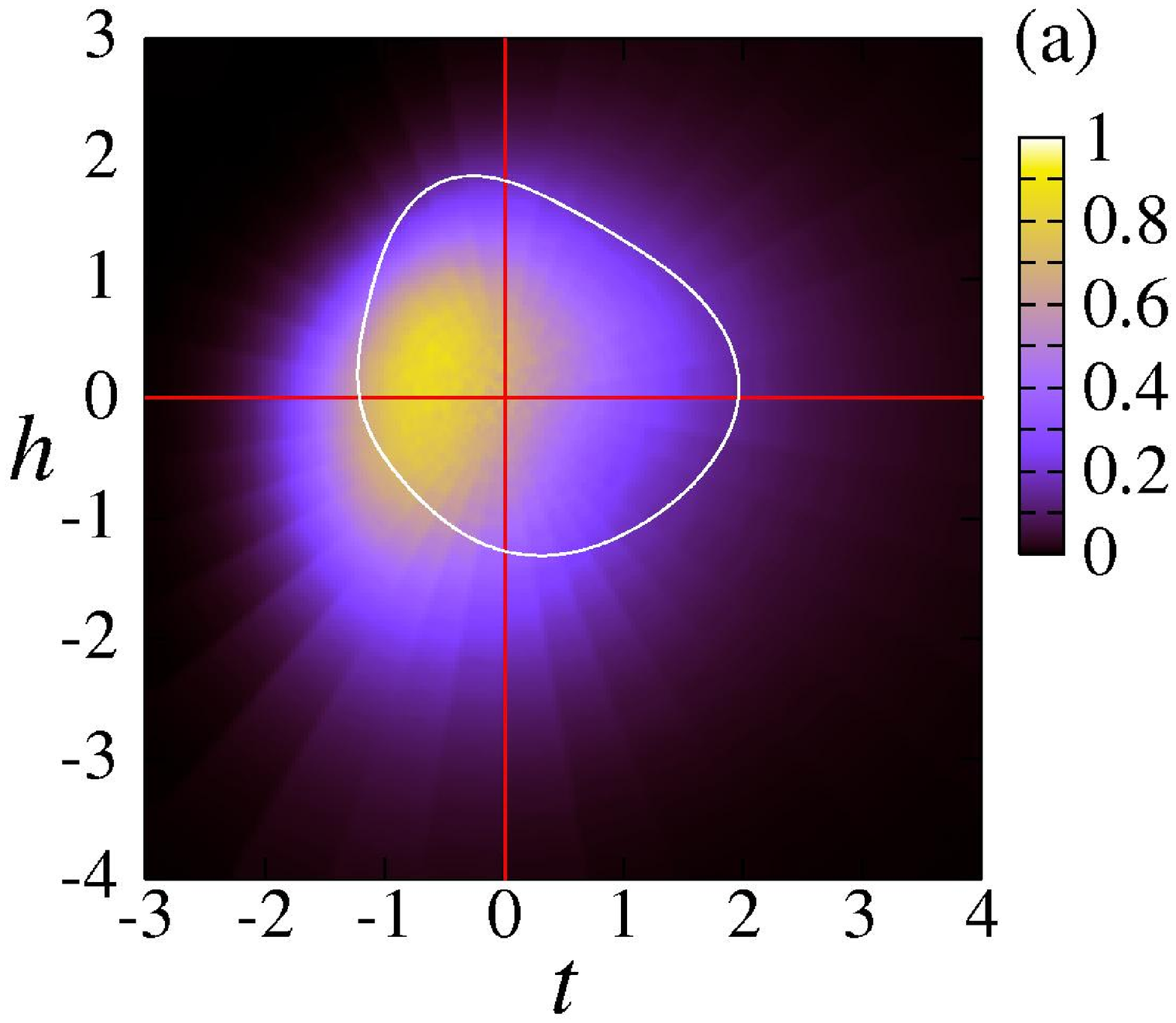}
\end{minipage}
\hspace{-5mm}
\begin{minipage}{0.49\linewidth}
\includegraphics [width=0.76\linewidth,angle=270]{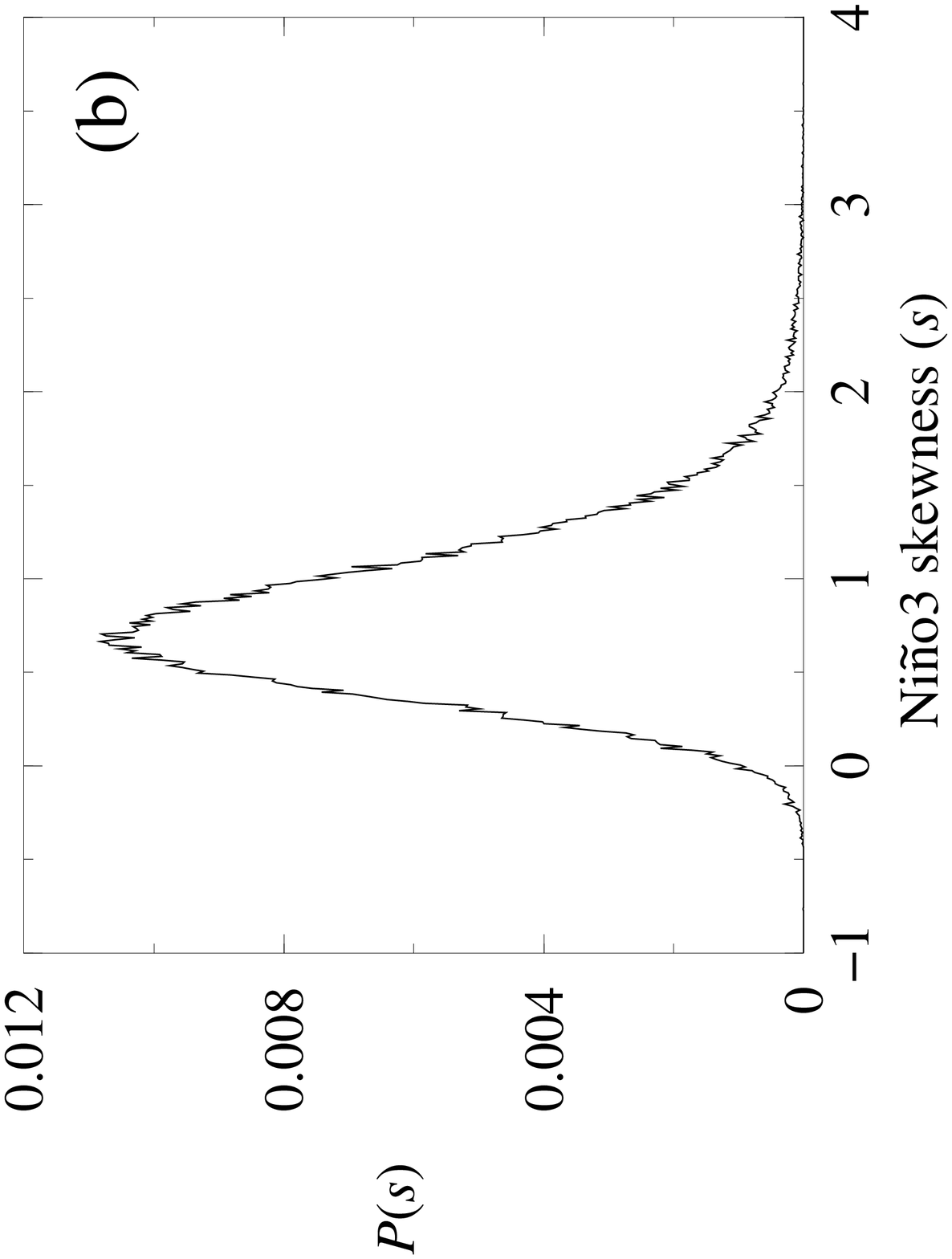}
\end{minipage}
\caption{(a): Probability density for finding an ENSO state at $(t,h)$
in colour plot. The white closed curve is the average path (obtained
by averaging the locations of all data points within
$10^\circ\,\phi$-intervals for the $5,000,000$ year run) of ENSO as a
stochastic limit cycle. It implies that on average an initial ENSO
state inside the white curve evolves in time to converge to it from
inside, while an initial ENSO state outside the white curve evolves in
time to converge to it from outside. The resemblance of
Fig. \ref{fig3}(a) to Fig. \ref{fig1}(a) or (b) is clear. Note the
relation of the white curve to the WWVA anomaly at the onset of an El
Ni\~no: a sharp positive anomaly in Equatorial WWVA precedes an El
Ni\~no, which is nicely captured by the hump in the white curve in the
$(t<0,h>0)$ quadrant. (b) Probability density $P(s)$ of skewness
$s$ of the Ni\~no3 index, calculated from $200,000$ sets of $25$ years
each (obtained from the same $5,000,000$ year run). The Ni\~no3
skewness based on the $25$ year UK MET office objective analysis data
(1980-2004) appearing in Fig. \ref{fig1}(a) is approximately $0.85$,
while the average Ni\~no3 skewness obtained from Fig. \ref{fig3}(b) is
$0.81$. The broad probability distribution of the Ni\~no3 skewness
exemplifies the extent of ENSO variability within the scope of this
model. These plots use the same values for the parameters as in
Fig. \ref{fig2}.\label {fig3} }
\end{center}
\end{figure}
The first issue for describing ENSO by means of these $16$-parameter
functional forms for $\gamma(r,\phi)$ and $\omega(r,\phi)$ is the
``stability'' of the optimization procedure; namely that all the major
ENSO features: irregularity, variability and Ni\~no3 skewness are
expected to be properly reproduced when the same minimization
procedure is applied to a dataset reasonably similar to the ABOM
one. The noise characteristics emerging from that new dataset are also
expected to appear to be sufficiently close to Gaussian white. To this
end the ENSEMBLES data [Fig. \ref{fig1}(a)] were used, and both
expectations were fulfilled. Not only did the phase-space manifold
structure of ENSO dynamics show clear signatures of growth and decay
in the ENSO magnitude $r$ [see Figs. \ref{fig2}(a) and (b)], but also
when the corresponding parameters were used, in addition to Gaussian
white noise (of strength obtained from the minimization procedure; see
methods section), to simulate the two-dimensional stochastic dynamical
system version of ENSO on a computer for $5,000,000$ years, it
produced the right 1980-2004 ENSO features. Two of these are shown in
Fig. \ref{fig3}. In Fig. \ref{fig3}(a) appears the probability density
function for finding the ENSO state at $(t,h)$ in colour plot. The
white closed curve is the average path (obtained by averaging the
locations of all data points within $10^\circ\,\phi$-intervals for the
$5,000,000$ year run) of ENSO as a stochastic limit cycle. It implies
that on average an initial ENSO state inside the white curve evolves
in time to converge to it from inside, while an initial ENSO state
outside the white curve evolves in time to converge to it from
outside. The resemblance of Fig. \ref{fig3}(a) to Fig. \ref{fig1}(a)
or (b), which prompted the ansatz to model ENSO as a limit cycle in
the first place, is clear. Also clear is the relation of the white
curve to the WWVA anomaly at the onset of an El Ni\~no: a sharp
positive anomaly in Equatorial WWVA precedes an El
Ni\~no\cite{meinen2,wang1,wang2}, which is nicely captured by the hump
in the white curve in the $(t<0,h>0)$ quadrant. Simultaneously,
Fig. \ref{fig3}(b) shows the probability density $P(s)$ of 
skewness $s$ of the Ni\~no3 index, calculated from $200,000$ sets of
$25$ years each (obtained from the same $5,000,000$ year run). The
skewness of the Ni\~no3 index based on the $25$ year (1980-2004)
ENSEMBLES data appearing in Fig. \ref{fig1}(a) is approximately
$0.85$, while the average Ni\~no3 skewness obtained from
Fig. \ref{fig3}(b) is $0.81$. The broad probability distribution of
the Ni\~no3 skewness exemplifies the extent of ENSO variability within
the scope of this model.

\section*{Predictability skill}

The ultimate test of how robustly ENSO can be described by a
two-dimensional stochastic dynamical system as above is to be able to
predict ENSO with a reasonable accuracy. In order to address this
issue, predictability for Ni\~no3 over the period $1960$-now was
studied by running a 10,000 different sequences of Gaussian white
noise realizations in this model, using the ENSEMBLES data (available
up to $1960$, roughly when records-keeping of reliable regular
subsurface Pacific temperature profile began). It is to be emphasized
here that (a) The use of subsurface Equatorial Pacific data to analyze
the relation between WWVA and ENSO has so far dated back to
$1980$. This study, therefore, is the first one to extend that to
pre-$1980$; (b) This model's predictability for Ni\~no3 has been
studied thoroughly by considering a wide variety of ``training
periods'', i.e., the periods from which the data are used to optimize
the parameters of the model. The results presented here are based on
the training period 1980-2004 as described below; this choice is
motivated as it also sheds light on the role of climate shifts on
ENSO; and (c) Unless otherwise stated, in order to avoid artificial
skill, the target period has {\bf never} been included in the training
period.
\begin{figure}
\begin{center}
\includegraphics [width=0.95\linewidth]{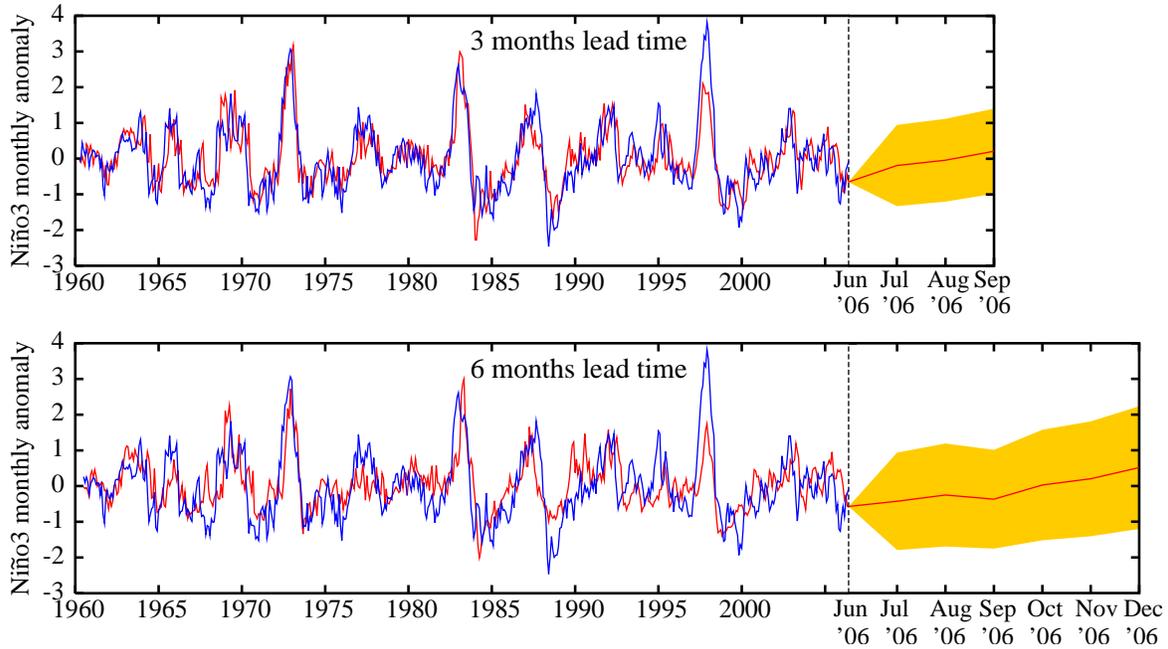}
\caption{Comparison between the predicted average (red) and the
observed (blue) values of normalized Ni\~no3 monthly anomalies over
the period $1960$-now (and beyond): lead time $3$ months (top) and $6$
months (bottom). The yellow areas show the $2\sigma$-spread in
the $3$- and $6$-month lead time predictions (they also give a general
idea about the magnitude of the $2\sigma$-spread over the $1960$-June
$2006$ period), based upon runs of 10,000 realizations. See methods
section for details.\label{fig4}}
\end{center}
\end{figure}

The predictability skill results (see methods section for the
calculations and the convention for lead time) for the years
$1960$-now appear in Fig. \ref{fig4}. The underlying subtleties can be
broken down to three separate periods. (i) $1980$-$2004$: In order to
make sure that the training period of the model did not include any
information from the ``target dataset'', i.e., the dataset
corresponding to the year for which Ni\~no3 is to be predicted, the
so-called ``jackknifing'' procedure was used. More explicitly, the
$(t,h)$ normalized anomaly dataset was first formed out of the
$1979$-$2004$ Ni\~no3 and Z20 monthly averages. Next, for the
predictions of year $n$ ($1980\leq n\leq2004$), the data for the years
$n-1$, $n$ and $n+1$ were removed from the $(t,h)$ dataset of
1979-2004 to form a jackknived dataset, which served as the training
period for the model. (ii) $1960$-$1979$: Two separate normalized
anomaly datasets $(t,h)$ were formed, one out of the Ni\~no3 and Z20
monthly averages of $1960$-$2004$ data, and the other out of the
$1980$-$2004$ data. The target dataset was formed by truncating the
former to $1960$-$1979$, while the training period for the model was
$1980$-$2004$. The comparison between the corresponding predicted and
observed $t$-values for the years $1960$-$1979$ revealed a startling
gap between the two (see Fig. \ref{appendd.1} in Appendix D), as if the
predicted magnitudes of Ni\~no3 were uniformly shifted upwards by
approximately $1^\circ$C compared to the observed ones over the period
1960-1976. Interestingly, when the Ni\~no3 and Z20 anomaly values for
$1960$-$2004$ were first passed through a high-pass filter that
removed the variations in these variables at time-scales $\geq 10$
years, and were subsequently normalized and truncated to $1960$-$1980$
in order to form the target dataset, the uniform gap between the
predicted and observed Ni\~no3 values disappeared (this is the
comparison shown in Fig. \ref{fig4}). Upon further reflection, it was
understood that this uniform gap corresponds to the well-known
climatological shift in $1976$ (known as the
``1976-shift''\cite{trenberth}). Interpreted differently, the
disappearance of the gap between predicted and filtered observation
data indicates that variations in the Equatorial Pacific climatic
conditions ENSO have two distinct components in it: climatological
shifts (i.e., shifts in the mean background) that occur due to
dynamics at decadal (or above) time-scales and variations that occur
at ENSO time-scales. It is the latter kind of variations that are
captured by the stochastic limit cycle picture. The issues related to
the origin of decadal variations in ENSO do remain matters of
discussion\cite{fedorov,flugel1,rodgers,yeh,schopf1}; nevertheless,
the conclusions of this analysis contradicts the view that
climatological shifts of ENSO are a part of its own dynamics. (iii)
$2005$-now (and beyond): The climatological means and normalizations
for the target dataset $(t,h)$ as well as the training period for the
model is $1980$-$2004$.

\begin{figure}
\begin{center}
\begin{minipage}{0.49\linewidth}
\includegraphics [width=0.75\linewidth,angle=270]{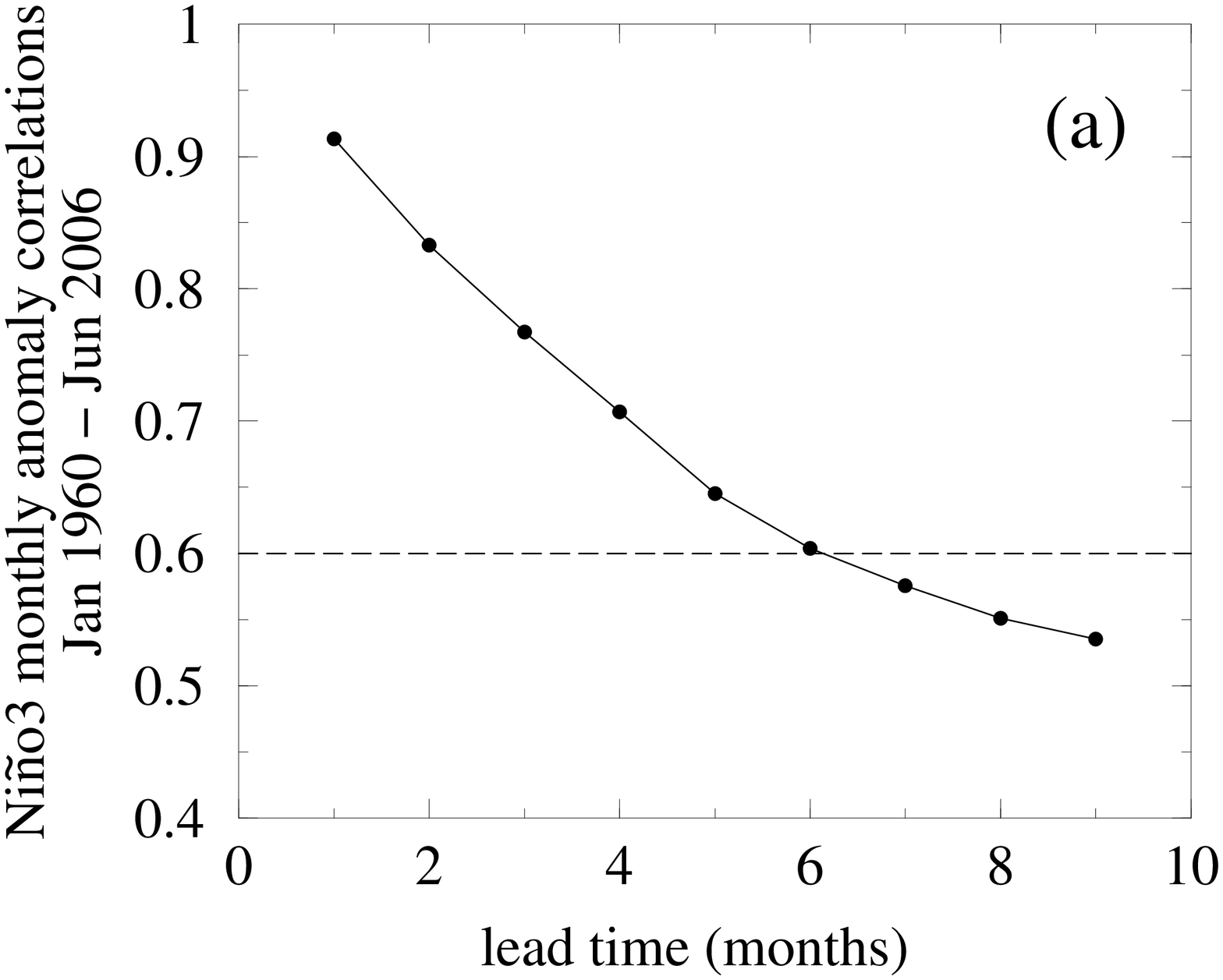}
\end{minipage}
\begin{minipage}{0.49\linewidth}
\includegraphics [width=0.75\linewidth,angle=270]{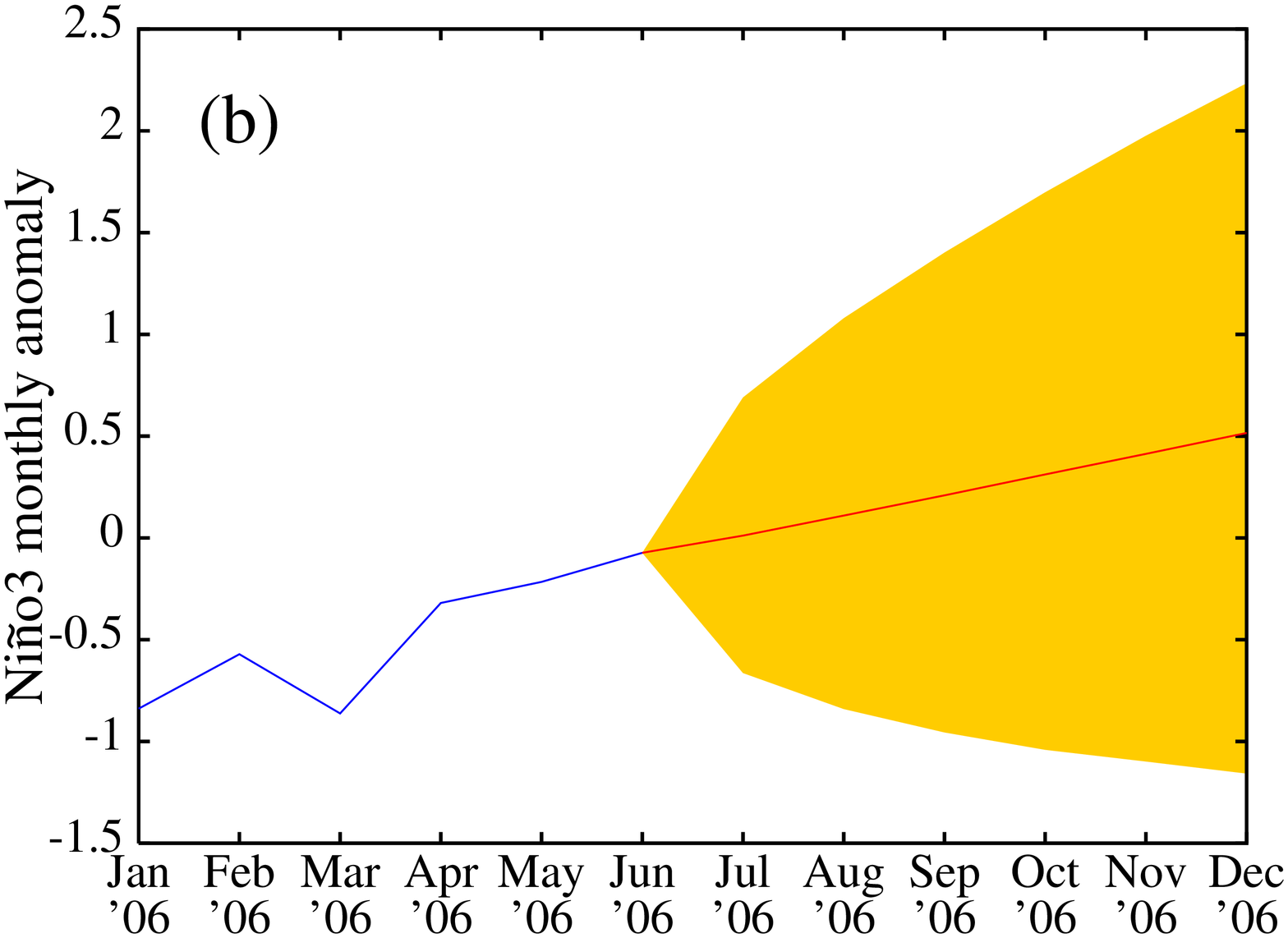}
\end{minipage}
\caption{(a) Predictability skill of this model for $1960$-June $2006$
(i.e., correlations between the red and blue curves of
Fig. \ref{fig4}). See also Fig. \ref{appendd.2} in Appendix D
for skill as a function of months. (b) Ni\~no3 monthly anomaly for 
2006, based upon the climatology of $1980$-$2004$; observation
(blue); predicted average (red). The yellow area shows the
$2\sigma$-spread. \label{fig5}}
\end{center}
\end{figure}
Finally, the predictability skill, i.e., Ni\~no3 monthly anomaly
correlations between observation and average of Ni\~no3 anomaly (as
they appear in Fig. \ref{fig4}) for $1960$-June 2006 are
shown in Fig. \ref{fig5}(a). The predictions until December 2006,
as obtained by (iii) above appear in Fig. \ref{fig5}(b).

\section*{Skill comparison with operational models} 

The comparison of the predictability skill of this model with the four
highest-ranked operational models [System-2 (S2)\cite{stockdale}
coupled atmosphere-ocean model from European meteorological centre
ECMWF, and Constructed Analogue (CA)\cite{dool1,dool2},
Markov\cite{xue} and Climate Forecast System (CFS)\cite{saha} from the
U.S. centre for meteorology NCEP] for the period $1987$-$2001$ are
presented in Table I. It provides a fair yardstick for the
predictability power of this model. Since different strategies were
used in different models to generate ensembles and the number of
members varied over the period $1987$-$2001$, making this comparison
has not been entirely trivial. Nevertheless, the period $1987$-$2001$
is actually selected as it is the common denominator of these
operational models\cite{gj}. See also Fig. \ref{appendd.3} in Appendix
D for skill comparisons as a function of months. 
\begin{center}
\begin{tabular}{||p{2.5cm}|p{5.7cm}|p{4cm}||}
\hline \hline 
& \hspace{2cm}lead months&\\
\hspace{1cm} model &
$\quad\,2\quad\quad\quad\quad\quad\,4\quad\quad\quad\quad\quad\,\,6$
&training/tuning period\\ 
\hline \hspace{1.5cm}S2&   
$0.94\quad\quad\quad\quad0.87\quad\quad\quad\quad0.76$ &
$\quad\quad\,\,1987$-$2001$\\ 
\hline \hspace{0.8cm}Markov &
$0.87\quad\quad\quad\quad0.75\quad\quad\quad\quad0.63$ &
$\quad\quad\,\,1981$-$1995$\\ \hline
\hspace{1.5cm}CA & 
$0.83\quad\quad\quad\quad0.76\quad\quad\quad\quad0.70$ &
$\quad\quad\,\,1956$-$2001$\\ \hline
\hspace{1.3cm}CFS &
$0.89\quad\quad\quad\quad0.80\quad\quad\quad\quad0.71$ &
$\quad\quad\,\,1987$-$2001$\\ \hline 
\hspace{0.5cm}This model & 
$0.90\quad\quad\quad\quad0.79\quad\quad\quad\quad0.69$ &
$\quad\quad\,\,1960$-$2004$\\
\hline \hline
\end{tabular}
\end{center}
Table I: Comparison of the prediction skill of this dynamical systems
model with the best operational models [S2 from ECMWF, and CA, Markov
and CFS from NCEP] for the period $1987$-$2001$. The CA and Markov
models are statistical, and the CFS and S2 models are coupled
atmosphere-ocean general circulation model (AO-GCM). Over this period,
the two-dimensional dynamical system model used here performs slightly
better than those of the NCEP statistical ones, and slightly worse
than the AO-GCM of ECMWF. Source: van Oldenborgh {\it et al.},
2003\cite{gj}.

It is also worth noting that a relatively recent work by Chen {\it et
al.}\cite{chen}, using training period $1980$-$2000$\cite{anderson}
reported that large El Ni\~nos  could be predicted up to two years in
advance. A predictability comparison with that work before $1960$ is
not possible because of the lack of Z20 data. Nevertheless, when the
training period $1980$-$2004$ is used, the model discussed in this
paper produces very similar skills on target periods $1960$-$1975$ and
on $1976$-$1995$ (Fig. \ref{appendd.4} in Appendix D).

\section*{Conclusion}

The idea behind viewing ENSO as a two-dimensional stochastic dynamical
system in this work has been based on the recharge oscillator
formulation. In this formulation, essentially the amount of warm water
in the Equatorial Pacific controls ENSO. Although it is clear that a
larger amount of warm water volume in the Equatorial Pacific results
in a higher eastern Pacific SST and vice versa, how these two
quantities interact is far from clear; a fact that this study draws
its motivation from. It considers subsurface Pacific data roughly
since record-keeping of reliable regular subsurface Pacific
temperature profile began (1960), for the first time. This simple
model firmly establishes that a two-dimensional dynamical system, and
more specifically, a nonlinear stochastic limit cycle captures the
essentials of ENSO to a great detail, as it (i) captures the
well-known features of ENSO comprehensively, (ii) is able to predict
ENSO with an accuracy well-comparable with the state-of-the-art
complex models from the European and American meteorological centres,
and (iii) is able to separate the nature of decadal and multi-decadal
variations from the interannual variability of ENSO. This development
signifies the fact that oceanic upwelling, vertical layer mixing, and
air-sea fluxes feedback mechanisms\cite{wang1,wang2} should be paid
more attention to in order to fully understand ENSO, as the
interaction between the subsurface warm water volume and the SST takes
place via these mechanisms. The expectation is that the parameters of
this dynamical system are related to these mechanisms, and that
remains a topic of future research.

\section*{Appendix A}

\subsection{Explanation for $\omega(r,\phi)\propto\sqrt{r}\,$:} 

This
proportionality has been actually chosen (i) to avoid the ambiguity of
defining $\omega(r,\phi)$ at $r=0$, (ii) as well as  for
``regularization'': should the system get stuck in a phase of
uncontrolled growth, the $\sqrt{r}$ term serves to move it quickly
away from there. In general, $\omega(r,\phi)\propto r^n$ for any $n>0$
can serve for both, so there is some arbitrariness in the choice of
$n$. Nevertheless, $n=1/2$ was a choice motivated to keep the model as
simple as possible: $n=0$ does not work for (i) and (ii), and $n=1$
was found to contradict the observation data, and hence the choice
$n=1/2$ was made. It is also worthwhile to note that the results
presented here are insensitive of the value of $n$ within the range
$0.3\lesssim n\lesssim0.8$.

\section*{Appendix B}

\subsection{Least-square optimization and the functional forms of
  $f_1(\phi)$, $f_2(\phi)$ and $g(\phi)\,$:} 

For given functional forms of $f_1(\phi)$, $f_2(\phi)$ and $g(\phi)$
parametrized by $n$ quantities (such as $A$, $B_1$ etc.), the
least-square optimization is a minimization of the scalar  function
$E$ defined on an $n$-dimensional manifold. In order to perform this
minimization the amoeba method was used\cite{press}. To make sure that
the true minimum for $E$ was reached, the minimization was started
from a multitude of initial values of the $n$ parameters. A single
minimization run on $25$ years of monthly data takes about 15 minutes
on a 1.8 GHz CPU. For the $1980$-$2002$ $(t,h)$ ABOM data initial trial
minimization runs were set up with
$f_1(\phi)=a_0+(b_{01}\cos\phi+c_{01}\sin\phi)+\ldots+(b_{04}\cos4\phi+c_{04}\sin4\phi)$,
$f_2(\phi)=1+b_1\cos\phi+c_1\sin\phi$, and
$g(\phi)=a_2+(b_{21}\cos\phi+c_{21}\sin\phi)+\ldots+(b_{24}\cos4\phi+c_{24}\sin4\phi)$.
It was found that increasing the number of parameters beyond $16$
hardly reduced the minimum value of $E$. Once the number of parameters
was thus fixed at $16$, more trials were run with different
combinations of sine and cosine harmonics. Eventually, for the ABOM data,
it was found that the forms of $f_1(\phi)$, $f_2(\phi)$ and $g(\phi)$
that lead to the minimum value of $E$ are the following:
$f_1(\phi)=a_0+b_{01}\cos\phi+b_{02}\cos2\phi+c_{02}\sin2\phi+b_{03}\cos3(\phi-\phi_{03})+b_{04}\cos4(\phi-\phi_{03}/2)$,
$f_2(\phi)=1+b_1\cos\phi+c_1\sin\phi$, and
$g(\phi)=a_2+b_{21}\cos\phi+c_{21}\sin\phi+b_{21}\cos2\phi+c_{21}\sin2\phi+b_{23}\cos3\phi+b_{24}\cos4\phi$.
These functional forms were used all throughout this study. The noise
properties of obtained from these data were found to be Gaussian white
to a good approximation. 

For later references, note here that when these functional forms were
used on the $1980$-$2004$ ENSEMBLES data [see Fig. \ref{fig1}(a)], the
noise characteristics again turned out Gaussian white to a good
approximation, with HWHM $0.34$ and $0.29$ for $t$ and $h$
respectively.

\section*{Appendix C}

\subsection{ENSO prediction as a function of lead months:}

The predictability studies for ENSO were performed in the usual way:
the numerical values of the model parameters were obtained from the
$(t, h)$ data over the training period. These parameters were then
used to time-evolve an initial state $(t_0, h_0)$ to $(t_i, h_i)$ for
$i=1,\ldots,12$ using $10,000$ different sequences of Gaussian white
noise realizations; this takes only about a minute on a 1.8 GHz
CPU. The HWHM used for the Gaussian white noise were $0.34$ for $t$ and
$0.29$ for $h$, as found from the optimization procedure on the
ENSEMBLES data. The predicted $(t_i, h_i)$ values constitute the
prediction for ``lead time $i$ months''.

\eject
\section*{Appendix D}

This appendix consists of additional figures to supplement the text.

\begin{figure}[!h]
\begin{center}
\includegraphics[width=0.55\linewidth,angle=270]{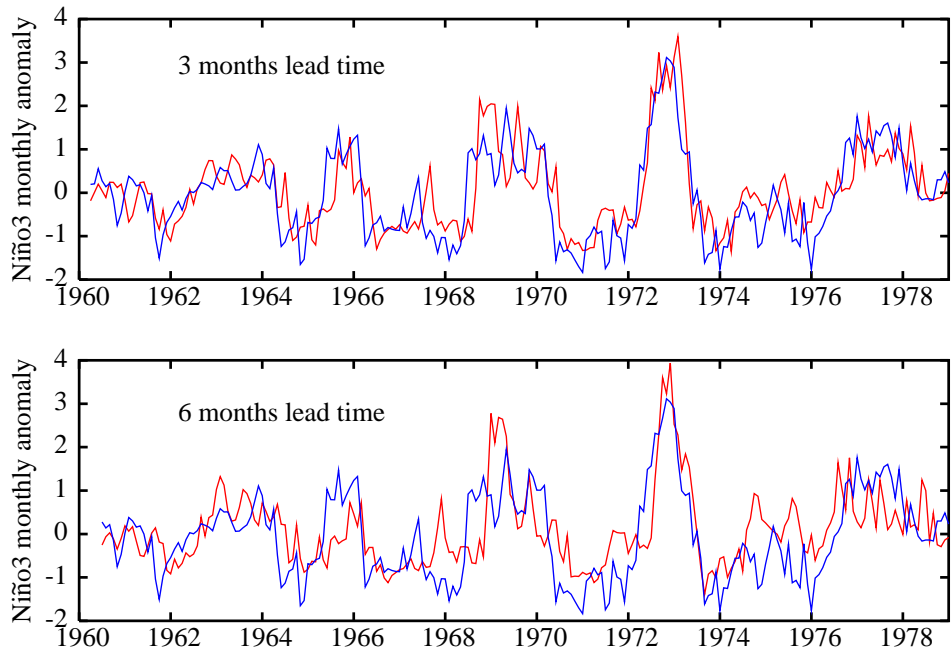}
\end{center}
\caption{Comparison between predicted (red) and unfiltered
observation data (blue) for Ni\~no3 monthly anomaly over
1960-1979. The uniform gap of nearly 1$^\circ$C between the two curves
before 1976 is due to the well-known 1976-shift. This gap disappears
(Fig. 4 in the paper) when the observation data is passed through a
high-pass filter that removes decadal (and above) variations in
Ni\~no3 and Z20.\label{appendd.1}}
\end{figure}
\eject
\begin{figure}[!h]
\begin{center}
\includegraphics[width=0.4\linewidth,angle=270]{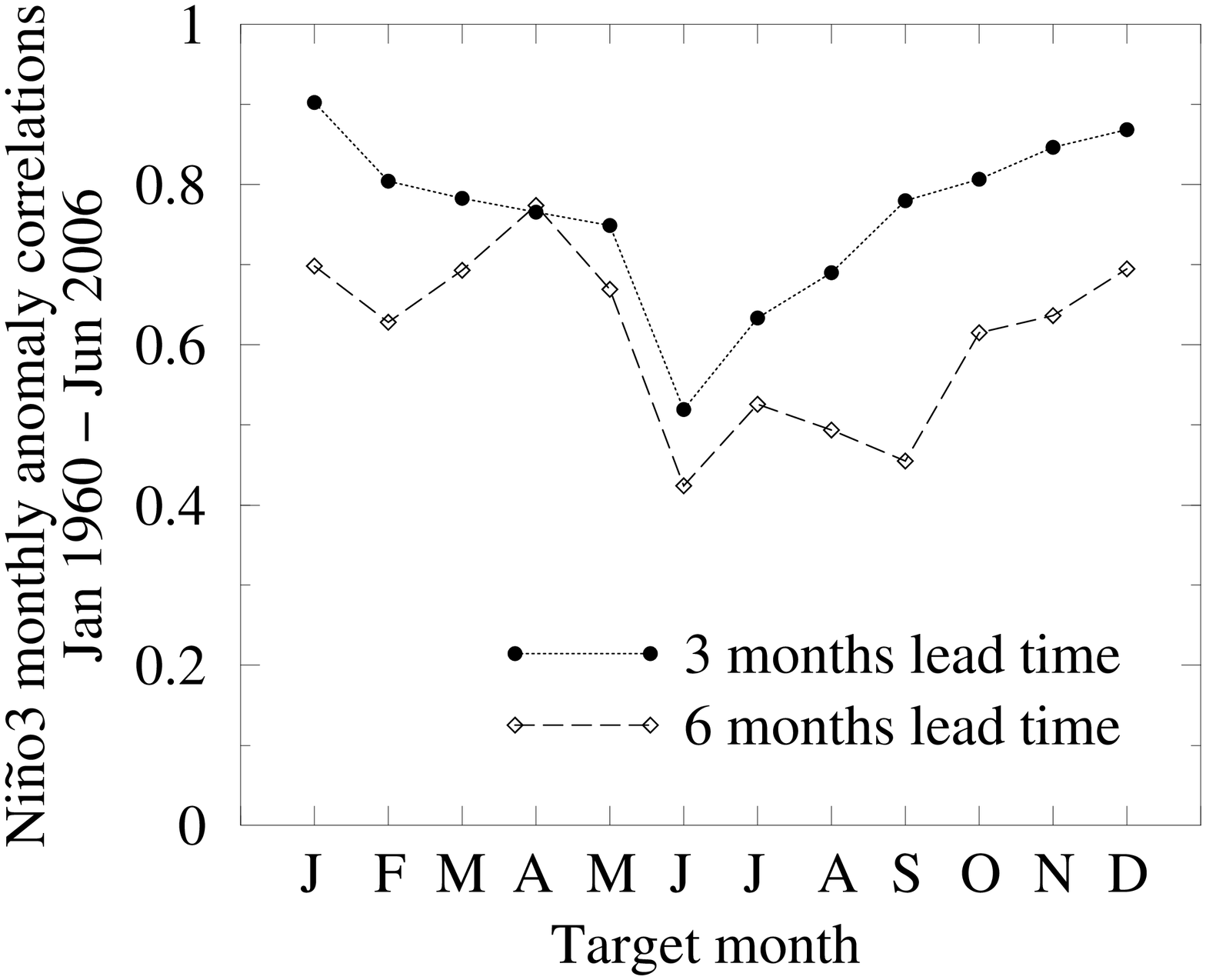}
\end{center}
\caption{Predictability skill of the dynamical system model as
a function  of target months for 1960-2006 (i.e., between the red and
blue curves of Fig. 4)\label{appendd.2}.}
\begin{center}
\includegraphics[width=0.4\linewidth,angle=270]{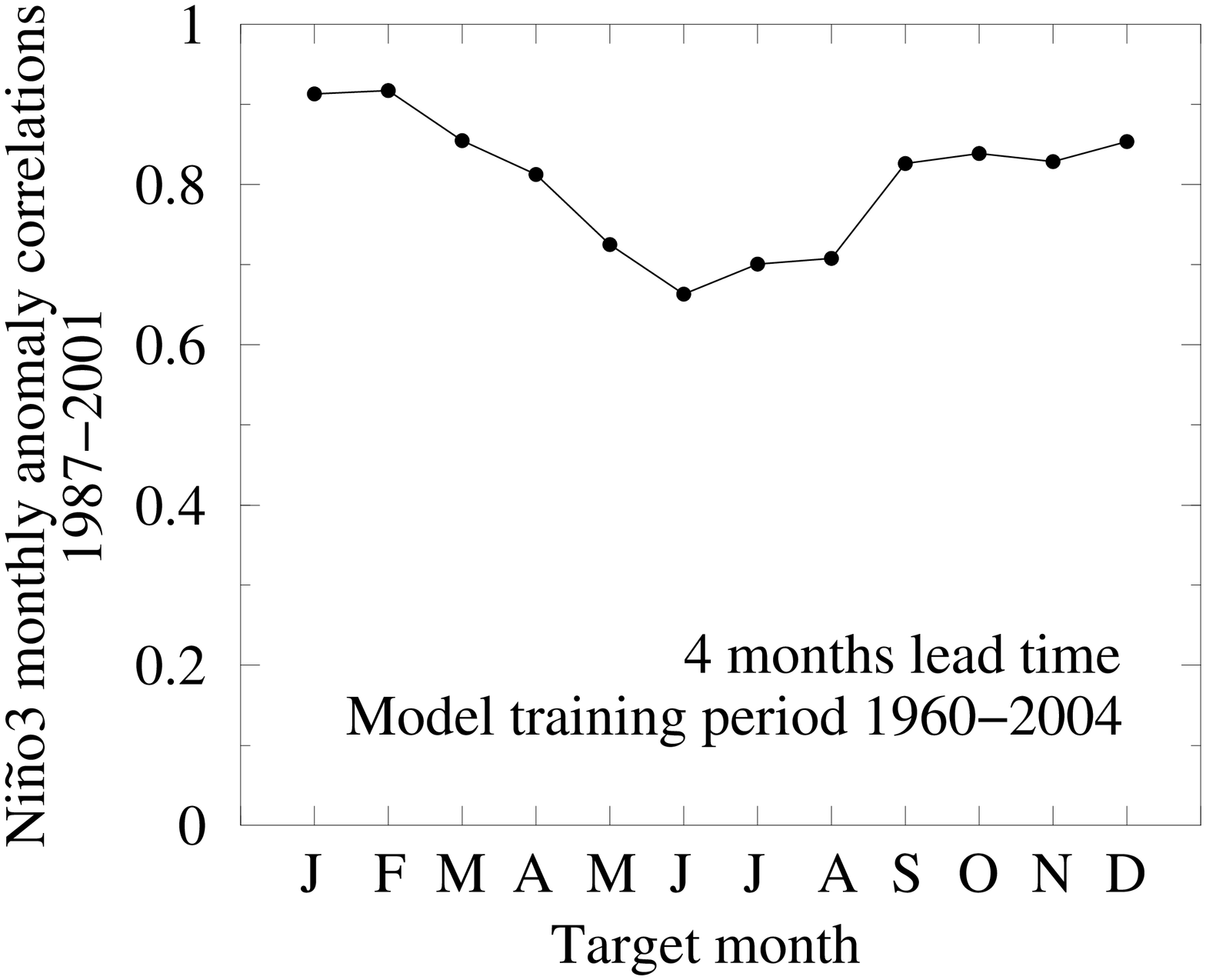}
\end{center}
\caption{Predictability skill of the dynamical system model in
the paper  as a function of target months. This figure is meant  for
comparison with Fig. 2 of van Oldenborgh {\it et al.} (2005),
which shows the ENSO  predictability skill for the operational models
of Table I.\label{appendd.3}}
\end{figure}

\begin{figure}[!t]
\begin{center}
\includegraphics[width=0.4\linewidth,angle=270]{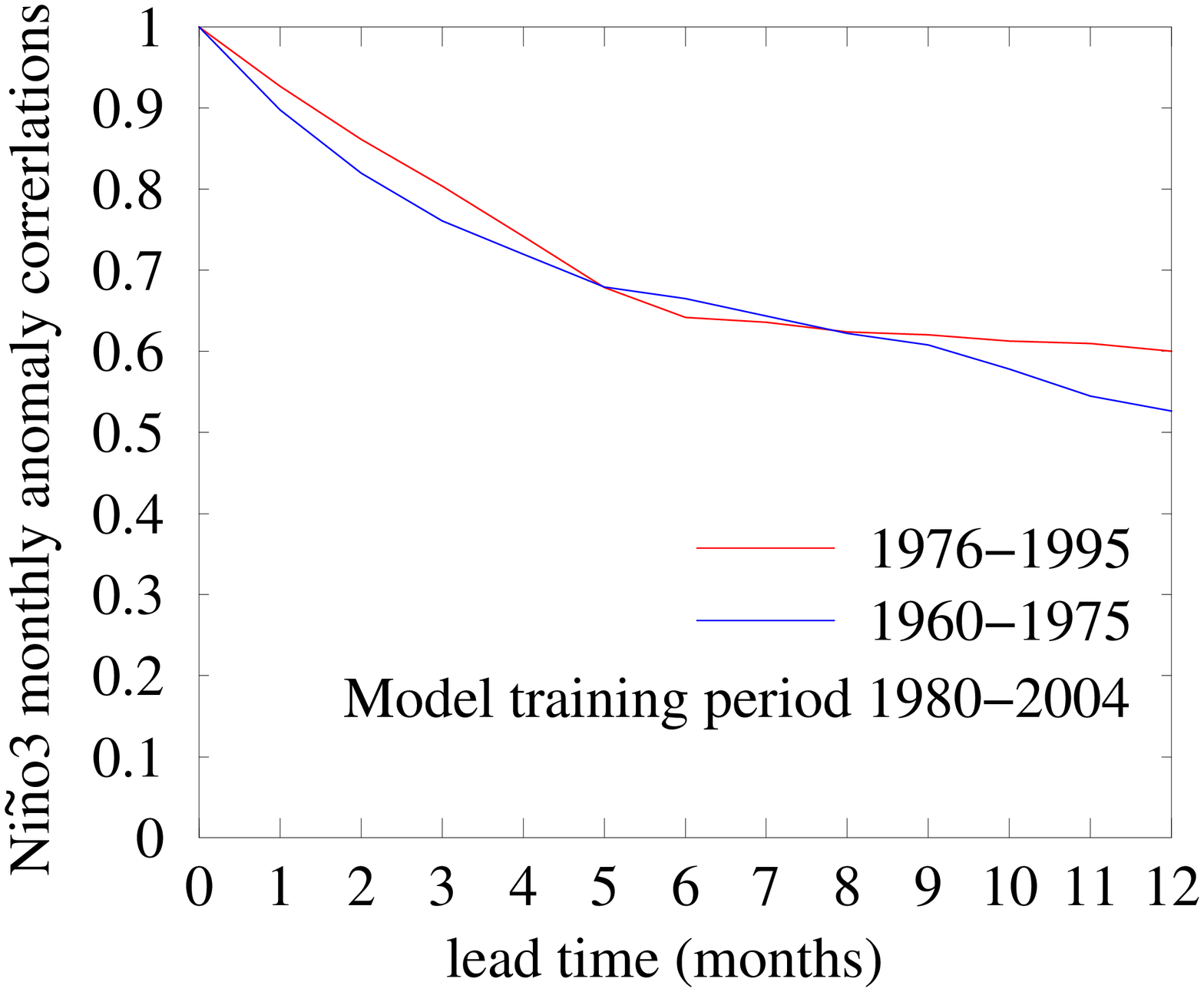}
\end{center}
\caption{Predictability skill of the dynamical systems model 
in the paper. This figure is meant for comparison with Fig. 2 of Chen
{\it et al.} (2004). To generate this figure, the decadal
(and above) variations have been removed from the 1960-1979 dataset, as 
discussed in the ``Predictability skill" section of the
paper.\label{appendd.4}}  
\end{figure}

\section*{References}


\bibliography{gran_stat}


\begin{addendum}
\item The authors wishe to thank Geert Jan van Oldenborgh, Wilco
Hazeleger and Ruben Pasmanter for useful discussions and comments on
the manuscript, and Bruce Ingleby at ECMWF and Sjoukje Philip for
their help with the data.
\end{addendum}

\end{document}